\documentclass[prd,aps,twocolumn,nofootinbib]{revtex4}

\usepackage{graphicx}

\begin{document}

\title{Formation of the $Q$ ball in the thermal logarithmic potential and its properties}

\author{Shinta Kasuya}

\affiliation{
Department of Information Sciences,
     Kanagawa University, Kanagawa 259-1293, Japan}

\date{February 22, 2010}

\begin{abstract}
We investigate the $Q$-ball formation in the thermal logarithmic potential by means of 
the lattice simulation, and reconfirm qualitatively the relation between $Q$-ball charge 
and the amplitude of the Affleck-Dine field at the onset of its oscillation. We find time 
dependence of some properties of the $Q$ ball, such as its size and the field value at 
its center. Since the thermal logarithmic potential decreases as the temperature falls down, 
the gravity-mediation potential will affect the properties of the $Q$ ball. Even in the case when
the gravity-mediation potential alone does not allow $Q$-ball solution, we find the 
transformation from the thick-wall type of the $Q$ ball to the thin-wall type,
contrary to the naive expectation that the $Q$ balls will be destroyed immediately 
when the gravity-mediation potential becomes dominant at the center of the $Q$ ball.
\end{abstract}


\maketitle


\section{Introduction}
$Q$-ball formation is ubiquitous in the Affleck-Dine mechanism for baryogenesis 
\cite{KuSh,EnMc,KK1,KK2,KK3}. Soon after the Affleck-Dine field begins rotation in the potential
which allows the $Q$-ball solution, the homogeneous field starts to fluctuate and transforms
into lumps. The actual formation was investigated numerically on the lattices for the gauge- and
gravity-mediated supersymmetry (SUSY) breaking scenarios \cite{KK1,KK2,KK3,Qsim}.

On the other hand, the properties of the $Q$ ball are quite different among the different forms 
of the potential. For example, for the flat potential, such as in the gauge-mediated SUSY
breaking scenario, the $Q$-ball size, the field value at its center, and the field rotation speed 
depend on the charge $Q$ of the $Q$ ball nontrivially \cite{DKS}.

The potential may be dominated by thermal effects after inflation. In the Affleck-Dine scenario,
the field amplitude at the onset of the rotation is very large, and the two-loop thermal effects on 
the potential are crucial \cite{AnDi,FHY}. This potential is given by
\begin{equation}
V_T \sim T^4 \log \left(\frac{|\Phi|^2}{T^2}\right),
\end{equation}
for large field values \footnote{
We do not consider a negative thermal logarithmic potential \cite{negalog,cg},
because $Q$-ball formation does not occur in that potential.}
. We called it the thermal logarithmic potential \cite{KK3}.
The energy density of the universe is dominated by the oscillation of the inflaton after inflation,
but there exists dilute plasma \cite{KT} which would build up the thermal logarithmic potential. 
We considered in Ref.~\cite{KK3} that the $Q$-ball formation in the thermal logarithmic potential
would be more or less similar to the time-independent logarithmic potential because of the 
fast growth of the fluctuations of the field in spite of the time dependence of the temperature $T$.
We thus borrowed the results from the time-independent logarithmic potential case. For example,
the charge of the produced $Q$ ball is given by \cite{KK3}
\begin{equation}
Q=\beta\left(\frac{|\Phi|}{T}\right)^4_{\rm osc},
\end{equation}
where $\beta\approx 6\times 10^{-4}$, and the subscript ``osc" denotes the values of the variables
at the onset of the oscillation of the field.

In this article we actually perform lattice simulations of the $Q$-ball formation in the thermal 
logarithmic potential, and see the evolution of the field and the distribution of the produced 
$Q$ balls. We also study of the properties of the formed $Q$ balls, especially about 
the evolution of the field amplitude at the center of the $Q$ ball.

In addition to the thermal logarithmic term, there would also be a mass term due to the 
gravity-mediated SUSY breaking effects for both the gauge- and gravity-mediation scenarios. 
The mass term alone (including one-loop potential) does not allow the $Q$-ball solution in 
some cases. In such cases, one may naively consider those $Q$ balls, created when the 
thermal logarithmic potential dominates, to disappear when the mass term begins to dominate 
the potential at the field value of the $Q$-ball center. 
As shown below, however, the $Q$-ball solution still exists later. Actually, the 
thick-wall type (gauge-mediation type) $Q$ ball transforms to the thin-wall type.

The structure of the article is as follows. In the next section, we show the results of lattice 
simulations for both time-independent and thermal logarithmic potentials. Some properties
of the $Q$ ball with thermal logarithmic potential are shown in Sec.III, while, in Sec. IV, we 
focus on the transition from the gauge-mediation type to the thin-wall type Q balls when 
the mass term in the potential gradually dominates over the thermal logarithmic potential. 
We finally conclude in Sec.V.

\section{Q-ball formation}
We investigate the $Q$-ball formation by means of three-dimensional lattice simulations.
Interpolating the thermal mass term at smaller field amplitudes and the thermal two-loop 
potential growing logarithmically at larger field values, we take the following form of the 
potential:
\begin{equation}
V_T(\Phi) = T^4 \log \left( 1+ \frac{|\Phi|^2}{T^2}\right).
\end{equation}
In addition, we restudy the time-independent case where $T$ is replaced by the constant 
mass $M_F$ for comparison, which is nothing but the potential of the gauge-mediated SUSY 
breaking effects. Here we consider the inflaton-oscillation dominated universe after inflation 
before reheating, but a similar argument applies also to the radiation dominated era. 
The temperature decreases as the universe expands as $T \propto a^{-3/8}\propto t^{-1/4}$, 
where $a(t)$ is the scale factor of the universe. Since we are interested in the period after the 
Affleck-Dine field starts the oscillation (rotation) when $H^2 \simeq |V''|$, we take initial 
conditions as\footnote{
It is implicitly assumed that the 
helical motion is dynamically achieved by so-called $A$ terms.}
\begin{eqnarray}
& & \varphi_1(0) = \varphi_0(1 + \delta_1), \quad \varphi'_1(0)=\delta_2, \nonumber \\
& & \varphi_2(0) = \delta_3, \quad \varphi'_2(0)=\sqrt{2}(1+\delta_4), \nonumber \\
& & \tau(0) = \frac{2}{3h} = \frac{\sqrt{2}}{3}\varphi_0,
\end{eqnarray}
where all the variables are normalized by the temperature at the onset of the oscillation, 
$T_{\rm osc}$, such that $\varphi=\Phi/\sqrt{2}T_{\rm osc}$, $h=H/T_{\rm osc}$, 
$\xi=T_{\rm osc} x$ and $\tau=T_{\rm osc} t$. Here we decompose the field into real 
and imaginary parts as $\varphi = \varphi_1+i\varphi_2$. $\delta$'s represent the 
fluctuations of $O(10^{-7})$. We mostly use $256^3$ lattices, but in order to see 
any box size effects we also perform on $350^3$ lattices in some cases, but find 
no crucial differences between them.

\begin{figure}[h]
\includegraphics[width=90mm]{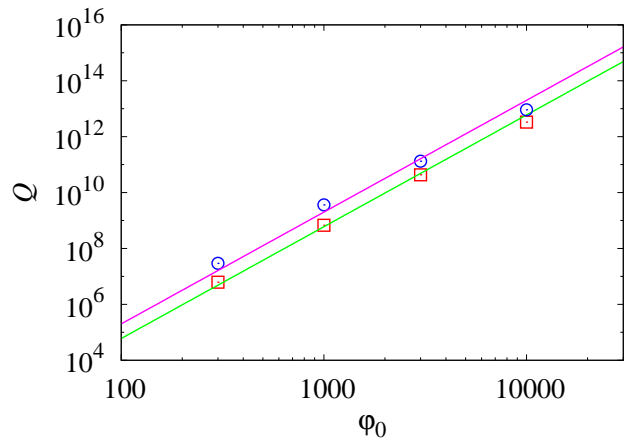} 
\caption{Charge of the Q ball depending on the amplitude of the Affleck-Dine field at 
the onset of the oscillation. 
The lower (upper) points and line correspond to the time-independent  (thermal) 
logarithmic potential.}
\label{fig_charge}
\end{figure}

Figure~\ref{fig_charge} shows the initial amplitude dependence of the largest charge 
of the $Q$ ball produced for the time-independent (lower) and thermal (upper) logarithmic 
potentials. Here we average the charge over five realizations of the initial fluctuations except 
for the smallest initial amplitude case ($\varphi_0=300$). The lower line corresponds to 
the relation obtained in Ref.~\cite{KK3},  $Q = \beta \varphi_0^4$ where 
$\beta\approx 6\times 10^{-4}$, thus we reconfirmed the previous results. 

On the other hand, in the thermal logarithmic potential, we have a similar relation
$Q=\beta' \varphi_0^4$ with $\beta'\approx 2\times 10^{-3}$, although it might have a little tilt.
Thus, qualitative features for the $Q$ balls in the thermal logarithmic potential can be
captured by the case with the time-independent logarithmic potential with $M_F$ being
replaced by $T_{osc}$.

\begin{figure}[h]
\includegraphics[width=85mm]{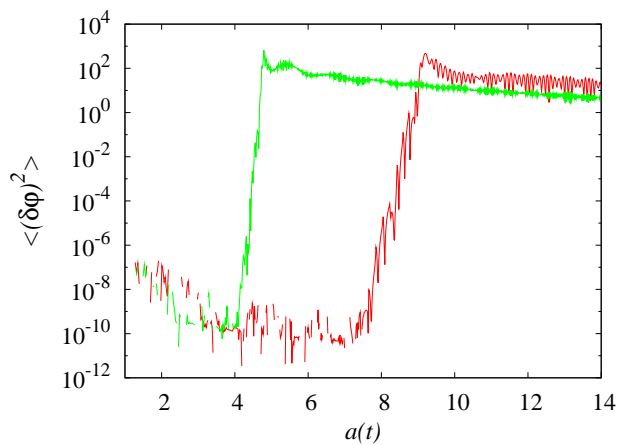} 
\caption{Evolution of the fluctuations of the Affleck-Dine in the time-independent (green, 
the earlier rise) and thermal (red, the later rise) logarithmic potentials for $\varphi_0=10^3$.
Notice that all the parameters including initial conditions are taken to be the same.}
\label{fig_fluct}
\end{figure}

The difference between the time-independent and thermal logarithmic potentials can be 
qualitatively considered as follows. $Q$-ball charge can be estimated as
$Q \sim q \ell_H^3$ at the formation time, where $q$ is the charge density and $\ell_H \sim t$ is
the horizon scale. From Fig.~\ref{fig_fluct}, the formation times are $a_0 \sim 4.6$ and 
$a_T \sim 9.3$, respectively, for the time-independent and thermal logarithmic potential. Thus, the 
ratio of the charges would be $\sim (a_T/a_0)^{-3}(t_T/t_0)^3\sim (a_T/a_0)^{3/2}\sim 3$.
Notice that the later rise and the slower growth of the amplitude of the fluctuations are due
to the shrinking instability band $0<k/a<2T^2/\phi_0$ and the lowering growth 
rate $T^2/(\sqrt{2}\phi_0)$ due to the decreasing temperature $T$.

The distribution of the $Q$ balls in example cases are shown in Table~\ref{dist-th} 
for the thermal logarithmic potential and in Table~\ref{dist-log} for the time-independent 
logarithmic potential, where we identified 
$Q$ balls with $Q > 10^3$ and $Q > 10^5$, respectively, in these cases. 
As one can see, the charge is dominated by those $Q$ balls with the charge of the 
largest magnitude. Therefore, it is fairly reasonable to estimate any relations among 
the $Q$-ball parameters by using the largest $Q$ ball, as we derived the charge 
and the initial field amplitude above. Since it is beyond the scope of the present paper 
to provide an analytical estimate of the distribution, we just provide a fitting formula of the form
\begin{equation}
N_i(\tilde{Q}) = \alpha_i \varphi_0 \tilde{Q}^{-\eta} 
e^{-\left(\frac{\tilde{Q}}{Q_{{\rm max}, i}}\right)^2}
\quad (i=0,T),
\end{equation}
where $\tilde{Q}$ denotes the charge in terms of the order of magnitude. Here 
$\eta\approx0.3$. $Q_{\rm max} = \beta_i \varphi_0^4$ with 
$\beta_0 \approx 6\times 10^{-4}$ and $\beta_T=2\times 10^{-3}$ for the 
time-independent and thermal logarithmic potentials, respectively. 
$\alpha_0 \approx 1.6$ and $\alpha_T\approx 1.1$.

\begin{table}[h]
\caption{\label{dist-th} Distribution of Q balls in the thermal logarithmic 
potential for $\varphi_0=10^3$.}
\begin{tabular}{ccccccc}
\hline\hline
Charge & $\ $ & Numbers & $\ $ & Sum of the charge & $\ $ & Fraction \\
\hline
$O(10^9)$ & & 2 & & $3.77 \times 10^9$ & & 0.7358 \\
$O(10^8)$ & & 2 & & $6.33 \times 10^8$ & & 0.1235 \\
$O(10^7)$ & & 15 & & $4.39 \times 10^8$ & & 0.0857 \\
$O(10^6)$ & & 15 & & $5.82 \times 10^7$ & & 0.0114 \\
$O(10^5)$ & & 12 & & $7.21 \times 10^6$ & & 0.0014 \\
$O(10^4)$ & & 51 & & $1.05 \times 10^6$ & & 0.0002 \\
$O(10^3)$ & & 195 & & $8.94 \times 10^6$ & & 0.0002 \\
\hline\hline
\end{tabular}
\end{table}

\begin{table}[h]
\caption{\label{dist-log} Distribution of Q balls in the time independent 
logarithmic potential for $\varphi_0=3\times 10^3$.}
\begin{tabular}{ccccccc}
\hline\hline
Charge & $\ $ & Numbers & $\ $ & Sum of the charge & $\ $ & Fraction \\
\hline
$O(10^{10})$ & & 5 & & $1.42 \times 10^{11}$ & & 0.7807 \\
$O(10^9)$ & & 10 & & $2.33 \times 10^{10}$ & & 0.1280 \\
$O(10^8)$ & & 20 & & $5.92 \times 10^9$ & & 0.0326 \\
$O(10^7)$ & & 44 & & $1.56 \times 10^9$ & & 0.0086 \\
$O(10^6)$ & & 56 & & $2.27 \times 10^8$ & & 0.0012 \\
$O(10^5)$ & & 163 & & $6.46 \times 10^7$ & & 0.0004 \\
\hline\hline
\end{tabular}
\end{table}

\section{Q-ball properties}
Let us investigate the evolution of the $D$-dimensional $Q$ ball formed in the thermal 
logarithmic potential, where for $D=1$ and 2 it is wall- and stringlike objects, respectively \cite{KK1}.
The charge of the $Q$ ball is given by
\begin{equation}
\label{constQ}
Q \sim a^3 R^D q \sim {\rm const.},
\end{equation}
where $q$ 
is the charge density and $R$ is the $Q$-ball size. Charge conservation
implies that $Q$ is constant. If we write $\Phi({\bf x},t) = \phi({\bf x})e^{i\omega t}/\sqrt{2}$,
the energy of the $Q$ ball is written as
\begin{eqnarray}
E & = & \int d^3x \left[\frac{1}{2}(\nabla\phi)^2+V(\phi)-\frac{1}{2}\omega^2\phi^2\right]+\omega Q
\nonumber \\
& = & \int d^3x \left[ E_{\rm grad} + V_1 + V_2 \right] +\omega Q,
\end{eqnarray}
where
\begin{eqnarray}
& & E_{\rm grad} \sim \frac{\phi^2}{a^2R^2}, \\
& & V_1 \sim T^4 \log\left( 1+\frac{\phi^2}{2T^2}\right) \sim T^4,\\
& & V_2 \sim \omega^2 \phi^2.
\end{eqnarray}
When the energy takes the minimum value, the equipartition is achieved for gauge-mediation type
$Q$ balls \cite{KK1}. From $E_{\rm grad} \sim V_1 \sim V_2$, together with the charge 
conservation (\ref{constQ}), we obtain the evolution of the (comoving) $Q$-ball size $R$, 
the rotation speed of the field $\omega$, and the field amplitude at the center of the 
$Q$ ball $\phi_c$, respectively as
\begin{eqnarray}
R & \propto & a^{-(4-\gamma)/(D+1)},\\
\omega & \propto & a^{-(D-3+\gamma)/(D+1)},\\
\phi_c & \propto & a^{(D-3-\frac{D-1}{2}\gamma)/(D+1)},
\end{eqnarray}
where we define $\gamma$ by $V_1 \sim T^4 \propto a^{-\gamma}$. Notice that 
$\gamma=3/2$ and 4 for the inflaton-oscillation and radiation dominated universe, respectively.
These properties are observed in the lattice simulations which we perform for the $D=3$ case.

\section{Transformation of Q-ball types}
In addition to the thermal logarithmic term in the potential, there is a mass term which stems 
from the gravity-mediated SUSY breaking effects. This potential can be written as
\begin{equation}
V_m = m_{3/2}^2 |\Phi|^2 \left[ 1+ K \log\left(\frac{|\Phi|^2}{M_*^2}\right)\right],
\end{equation}
where $m_{3/2}$ is the gravitino mass, and one-loop effects are included. $K$ is either
a positive or negative constant of $O(0.01 -0.1)$, and $M_*$ is a normalization scale.
This potential alone allows the $Q$-ball solution only if $K<0$ \cite{EnMc,New}. 
In the opposite case ($K>0$), $Q$-ball formation is prohibited. One may thus be apt to consider 
that the $Q$ ball created in the thermal logarithmic potential will disappear once the mass 
term with $K>0$ dominates over the thermal one at the field value of the $Q$-ball center, 
$V_T(\phi_c) < V_m(\phi_c)$. This condition can be written as
\begin{equation}
\phi_c > \phi_{eq}\sim  \frac{T^2}{m_{3/2}}.
\end{equation}
Since $\phi_{eq} \propto a^{-3/4}$ and $\phi_c \propto a^{-3/8}$ for $D=3$ in the 
inflaton-oscillation domination, and $\phi_{eq} \propto a^{-2}$ and $\phi_c \propto a^{-1}$ 
for any $D$ in the radiation domination, it is true that $V_m(\phi_c)$ will eventually 
overcome $V_T(\phi_c)$. 

However, the $Q$-ball solution {\it does} exist for $V=V_T+V_m$ with $K>0$ for
$T \gtrsim m_{3/2}$. Therefore, $Q$ balls are not destroyed, but the metamorphosis will 
take place in such situations. It might be best to simulate on the lattices to verify
this phenomenon, but it is very time-consuming to perform. Here, instead, we take
another approach, and leave the lattice simulations for future work.

We seek the $Q$-ball solution for $V=V_T+V_m$ at some time snapshots. In order to
obtain the solution, we just have to solve the equation
\begin{equation}
\frac{d^2\phi}{dr^2}+\frac{2}{r}\frac{d\phi}{dr}+\left(\omega^2\phi-\frac{dV}{d\phi}\right)=0,
\end{equation}
with boundary conditions $\phi(\infty)=0$ and $\phi'(0)=0$ \cite{small,KK4}.

\begin{figure}[ht]
\includegraphics[width=90mm]{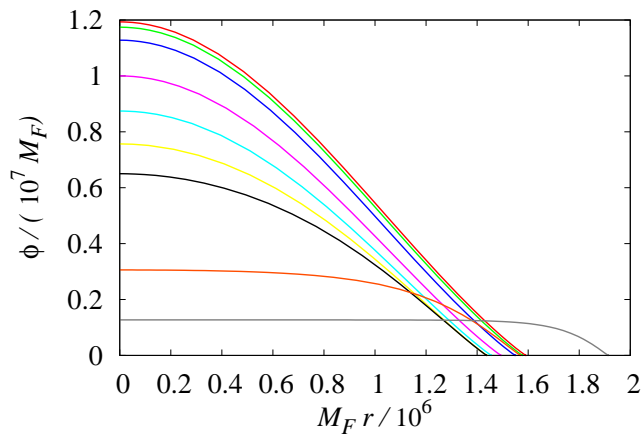} 
\caption{Profiles of the Q balls with $Q\approx 7.1 \times 10^{26}$ for 
$m/M_F = (0, 0.5, 1, 2, 3, 4, 5, 10, 20) \times 10^{-7}$ from the top to the bottom.
The $Q$-ball profile ``evolves" from the top to the bottom.}
\label{fig_prof}
\end{figure}

Since we would like to compare the results also to the solution of the pure logarithmic 
potential case, we change the value of the mass $m_{3/2}$ in $V_m$, leaving the 
logarithmic potential time independent. Thus we use the following potential with 
various values for $m_{3/2}$:
\begin{equation}
V=M_F^4\log\left(1+\frac{\phi^2}{2M_F^2}\right)
+\frac{1}{2} m_{3/2}^2 \phi^2\left(1+K\log\frac{\phi^2}{2M_*^2}\right).
\end{equation}
Notice that it is practically the same if one varies $M_F$ while $m_{3/2}$ is fixed, since
the results are derived and shown by variables normalized with respect to $M_F$.
The profiles are shown in Fig.~\ref{fig_prof} for 
$m_{3/2}/M_F= (0, 0.5, 1, 2, 3, 4, 5, 10, 20) \times 10^{-7}$ from the top to the bottom.
Here we set $K=0.1$ and $M_*/M_F=10^6$. It mimics the time evolution that the mass 
term eventually dominates over the logarithmic potential. Since the charge should be 
conserved, the angular velocity $\omega$ increases: 
$\omega/M_F=(2.0, 2.1, 2.4, 3.2,  4.1, 5.2, 6.2, 11, 21)\times 10^{-6}$ from the top to the
bottom. One can see that the thick-wall type (gauge-mediation type) of $Q$ ball 
transforms into the thin-wall-like type as time goes on. Notice that $\phi_c > \phi_{eq}$ 
takes place for $m/M_F \gtrsim 10^{-7}$.

In the gravity-mediation, $Q$-ball solutions may exist until $T \sim m_{3/2}$ for large 
enough $\omega$.~\footnote{
The fate of the $Q$ ball will depend on the cosmological situation. For thorough analysis,
see Ref.~\cite{CKY}.}
After that, $Q$ balls will disappear quickly, but the Affleck-Dine field 
can no longer be regarded as homogeneous because the field is localized near the place 
where destroyed $Q$ balls had existed; they were very much separated from each other. 
Typical separation length is estimated as 
$\ell_{H,{\rm formation}} (a_{\rm destruction}/a_{\rm formation})$. On the other hand,
in the gauge-mediation, $Q$ balls will remain intact for $m_\phi > m_{3/2}$, where 
$m_\phi=\sqrt{V''(0)}$ is the curvature of the gauge-mediation potential at $\phi=0$, 
since the $Q$-ball solution will exist irrespective of the temperature.

\section{Conclusions}
We have investigated the $Q$-ball formation in the thermal logarithmic potential
by means of three dimensional lattice simulations. First of all, $Q$ balls are actually formed.
This is because the growth of the field fluctuations is fast enough to create $Q$ balls,
in spite of the shrinking instability band due to the decreasing temperature, and so on. 
We have found that the charge of the $Q$ ball in the thermal logarithmic potential has 
almost the same dependence on the initial amplitude of the Affleck-Dine field,
\begin{equation}
Q=\beta' \left(\frac{\phi_0}{T_{\rm init}}\right)^4,
\end{equation}
with $\beta \approx 2 \times 10^{-3}$, which is a factor of 3 larger than that in the
time-independent logarithmic potential case.

We have also estimated the evolutions of parameters, such as the $Q$-ball size $R$, 
the field rotation velocity $\omega$, and the field value at the center of the $Q$ ball $\phi_c$.
Since the thermal logarithmic potential decreases as the temperature drops, the mass 
term would dominate over the thermal logarithmic one at the field value $\phi=\phi_c$.
Even if the mass term $V_m$ with positive $K$ alone does not allow any $Q$-ball solution,
the total potential $V=V_T+V_m$ {\it does} allow the solution. In fact, we have found that the 
thick-wall type Q ball will eventually transform into the thin-wall type. 
Finally, the $Q$ balls will disappear when $T \sim m_{3/2}$, with an 
inhomogeneous Affleck-Dine field being left afterward in the gravity-mediation scenario.

\section*{Acknowledgments}
The author is grateful to Masahide Yamaguchi for useful discussion.



\end{document}